# Spectral tunability of a plasmonic antenna with a dielectric nanocrystal

Yury Alaverdyan*, Nick Vamivakas, Joshua Barnes, Claire Lebouteiller, Jack Hare and Mete Atatüre*

*Cavendish Laboratory, University of Cambridge, JJ Thomson Avenue, Cambridge CB3 0HE, United Kingdom*
*ma424@cam.ac.uk, ya236@cam.ac.uk

**Abstract.** We show that the positioning of a nanometer length scale dielectric object, such as a diamond nanocrystal, in the vicinity of a gold bowtie nanoantenna can be used to tune the plasmonic mode spectrum on the order of a linewidth. We further show that the intrinsic luminescence of gold enhanced in the presence of nanometer-scale roughness couples efficiently to the plasmon mode and carries the same polarization anisotropy. Our findings have direct implications for cavity quantum electrodynamics related applications of hybrid antenna-emitter complexes.

## 1. Introduction

Collective oscillations of conduction band electrons confined at metal-dielectric interfaces give rise to surface plasmon polariton modes, which can confine electromagnetic fields to an effective volume that is considerably smaller than the optical diffraction limit. The field of plasmonics gained increasing attention following the discovery of surface-enhanced Raman scattering [1, 2] and development of surface plasmon resonance sensing schemes [3-7]. Recently, localized surface plasmon (LSP) modes have been considered as promising candidates for the strong light confinement needed in cavity quantum electrodynamics to achieve suppression of spontaneous emission, single photon generation in a well-defined spatial mode, and local probing of the environment [8-15]. For this purpose the bowtie (BT) antenna, one of the smallest plasmonic structures, simultaneously offers both extraordinary field confinement and a broadband spectral response [13-16]. The structure comprises two metal triangles with tips pointing towards each other separated by a small gap. The spatial distribution of the fundamental dipolar-like LSP mode for the polarization parallel to the antenna axis is suitable for coupling to an optical transition of a single emitter.

In this Letter, we show how the fundamental plasmon mode of a BT antenna can be tuned spectrally by controlling the position of a single diamond nanocrystal (NC) in the proximity of the antenna gap. Position control is achieved with an atomic force microscope in contact mode. For each crystal position, we measure dark-field scattering (DF) and antenna photoluminescence (PL) spectra of the BT-NC hybrid structure. We further report that the gold luminescence due to single electron intra-band excitation is dramatically enhanced due to efficient coupling to the modes of the BT antenna. In the fundamental dipolar-like mode, which is of interest for single emitter-cavity QED, this luminescence shows 150-fold enhancement.

## 2. Fabrication, simulations, manipulation and optical measurements

The BT antennas were fabricated using electron-beam lithography with positive resist ZEP520A on a 300 micron-thick quartz substrate. After the exposure (Crestec 9500C, 50 kV, 10 pA) and development in hexylacetate, 1 nm Cr and 29 nm Au were thermally deposited, followed by a lift-off in Shipley Remover 1165 at 60°C. Scanning electron micrographs (SEM) confirm that both the width and the length of each triangle in a BT antenna was 75 nm, with a gap of 25 nm (Fig. 1a). The diamond nanocrystals of 0-50 nm size range (Microdiamant) were dispersed in ethanol, ultra-sonicated for 1 hour, and then deposited on the BT-patterned substrate. The NC used for manipulations was ~35 nm in diameter. Scanning electron micrograph of a typical NC is shown in Fig 1b.

Finite difference time domain (FDTD) numerical simulations of the antenna-nanocrystal hybrid structure were performed using commercially available software (FDTD Solutions, Lumerical). BT corners were rounded to 9 nm radius to simulate the real structure (Fig. 1c) with experimentally determined spectral dependence of the refractive index for gold [17], chromium and quartz [18]. The NC is represented by two truncated pyramids one atop the other, each 17.5 nm in height (for a total 35 nm), 20 nm across at their tips and 35 nm across at their base, modelled as a dielectric with refractive index of 2.417 (Fig. 1d). Meshing was set at 0.5 nm in a region covering the BT plus crystal combination, and was progressively relaxed to automatically-determined larger cell sizes outside of this centre region. The boundary conditions were perfectly matched layers. In addition, the four simulations which are symmetric structures along the x axis (orthogonal to the bowtie axis), used an anti-symmetric minimum boundary condition for the x axis. The source is a plane wave, linearly polarised along the bowtie axis. The near-field profile images are taken at the substrate level and in the middle of the NC (17.5 nm above the substrate surface).

Optical measurements were performed with a home-built fibre-based confocal optical scanning microscope, shown in Fig. 1e. A halogen lamp was used as a broadband light source for DF scattering on the BT antenna. A 532-nm laser (Verdi, Coherent) was used for the generation of the gold luminescence for PL measurements. Each time-integrated DF spectrum was divided by the original halogen source spectrum for normalization. Imaging of the sample's topography was done using an atomic force microscope (AFM) (NanoWizard II, JPK) in tapping mode, while manipulation was performed in contact mode.

## 3. Results and discussion

Figures 2a-d display the AFM images of the BT antenna and the NC for five separate positions of the NC, ca. 108, 40, 20 and 16 nm away from the centre of the antenna gap, while Fig. 2e shows the NC in contact with one of the outer corners of the BT antenna. The measurements refer to the distance from the centre of the NC to the centre of the antenna gap. Scattering spectra taken from 24 identically fabricated BT antennas (not shown) exhibited a resonance wavelength uncertainty of 20 nm due to fabrication irregularities. All results discussed here were performed on the same BT-NC system. The distances between the centre of the gap and the crystal exhibit a tip-curvature related lateral uncertainty of 10 nm.

The corresponding simulated near-field images for the same crystal positions relative to the BT are shown in Fig. 2f-m. When the NC is in the vicinity of the antenna gap, the field distribution of the fundamental plasmon mode is modified significantly at the height corresponding to the largest lateral extent of the NC (17.5 nm above the substrate). However, the field distribution in the antenna gap at the substrate level, where a single emitter can be placed, is almost unaffected by the NC, therefore the coupling strength of an emitter to the plasmon mode can be maintained in the presence of a nearby NC.

The shape and size of the NC is chosen so that the centre of the antenna gap is always available for a single quantum emitter such as a single molecule or a nitrogen-vacancy centre in a small (~5 nm) diamond nanocrystal to be positioned for plasmonic coupling. Such systems, however, are commonly excited optically at energies higher than the fundamental emission energy. The exposure of the BT antenna to the typical nitrogen-vacancy center excitation wavelength of 532 nm results in the generation of gold luminescence signal [19-21]. This signal originates from the intra-band light absorption in gold and the subsequent radiative decay probability is greatly enhanced in the presence of nanometer-scale roughness (as seen in thermally deposited polycrystalline gold) due to the relaxed dispersion relation at this length scale. This results in a broad gold-generated luminescence spectrum which overlaps with the tailored plasmon resonances in the visible to near-infrared spectrum.

Figure 3a presents luminescence spectra from the BT antenna and from an extended gold film, illuminated by a 532-nm laser. The black and gray curves present the broad luminescence spectrum from an extended gold film for two orthogonal polarization measurement axes. The luminescence is unpolarized since the extended gold film does not possess polarisation-selective plasmon resonances. The red and blue curves in Fig. 3a denote the BT antenna luminescence measurements in two polarization axes, which are parallel and orthogonal to the BT axis, displaying the polarization anisotropy of a BT antenna [22, 23]. The degree of polarization anisotropy and the strength of luminescence do not depend on the polarization of the excitation laser. Taking into account the ratio of the excitation areas for the extended gold film and the BT antenna (~12), we deduce a 150-fold enhancement of the gold luminescence in the fundamental plasmon mode at the peak wavelength in comparison to that expected from a section of the same area as the BT antenna, in an extended gold film. This is indicative of the plasmon-assisted enhancement of gold luminescence and the strong feeding of the plasmon mode. The modal selectivity of this luminescence enhancement is also visible in Fig. 3a, where the fundamental mode is more than an order magnitude stronger at its peak wavelength than the orthogonally polarized luminescence component. The feeding of the BT antenna due to the luminescence of physically deposited gold is detrimental to applications requiring cavity-emitter coupling. This renders the vacuum Rabi splitting, a spectral signature of coherent (strong) coupling, unobservable, especially when the emitter lifetime becomes comparable to the mode lifetime. Recent reports on chemically-grown monocrystalline gold flakes [24, 25] are quite promising for resolving this fundamental problem, and our measurements on similar gold flakes support the suppressed luminescence (not shown). In this work, we choose to focus on utilizing the plasmon-coupled gold luminescence as a method to measure the dependence of the plasmon mode spectrum on the dielectric environment.

Figure 3b presents normalised DF and PL spectra along with the simulated scattering spectra for the five positions of the NC, discussed in Fig. 2. Each spectrum is fit with a Voigt function to determine the peak wavelength. While the peak wavelength depends significantly on the NC distance to the centre of the antenna gap, the plasmon resonance lineshape is essentially unchanged by the presence of the NC. The FDTD simulations of the plasmon lineshape (top set of curves) confirm that the lineshape is unaffected by the presence of the NC. Figure 3c shows the dependence of the resonance wavelength on the NC distance to the gap centre. As the overlap between the NC and the near-field profile of the plasmon mode increases, the plasmon mode samples more of the higher index dielectric and the observed resonance exhibits a nonlinear shift to longer wavelengths. The measured dependence of the wavelength on the NC position shows reasonable agreement with our simulations. Finally we position the NC next to one of the outer corners of the BT antenna and measure the spectrum of the fundamental plasmon mode. Figure 3c shows that the peak wavelength for this configuration is nearly identical to that of the BT without the NC (Fig. 2f). Figure 3b also confirms that the plasmon mode lineshape is unaffected by the NC at the outer corner. Figure 2 panels f-i show that the field is still strong at the outer corners at the substrate level, while it is strongly reduced at 17.5 nm above the substrate (Figs. 2j-m). Therefore, strong spectral shift would be observed if the NC had overlap with the mode at the substrate level. The absence of any spectral shift is the experimental verification that the NC we are positioning has a smaller footprint at the substrate level than its waist and that the spectral tuning of the mode is achieved by accessing the field well above the substrate.

## 5. Conclusion

Here we show that a dielectric nanocrystal can be used for tuning the resonance wavelength of a plasmonic BT antenna on the order of a linewidth. This ability is essential for controlling the spectral overlap of a quantum emitter, such as a single molecule, a quantum dot, or a diamond color center, with a plasmonic cavity mode. The spectral tuning is achieved by modifying the mode sufficiently above the substrate level in our experiments, and therefore, the anticipated spatial overlap of a quantum emitter with the mode distribution would be unaffected at the substrate level. We further show that gold luminescence is enhanced by the fundamental plasmon mode of a BT antenna by more than two orders of magnitude. The degree of coupling of the gold luminescence to the BT antenna is a roadblock for applications requiring coherent (strong) emitter-cavity coupling, but it can be used as a tool for characterizing the spectral properties of gold-based plasmonic nanostructures.

## 6. Acknowledgements

We thank T. Müller for technical assistance. The research leading to these results has received funding from the European Research Council (FP7/2007-2013)/ERC Grant agreement No. 209636, the internal funds of the University of Cambridge and EPSRC.

**Figure 1.**

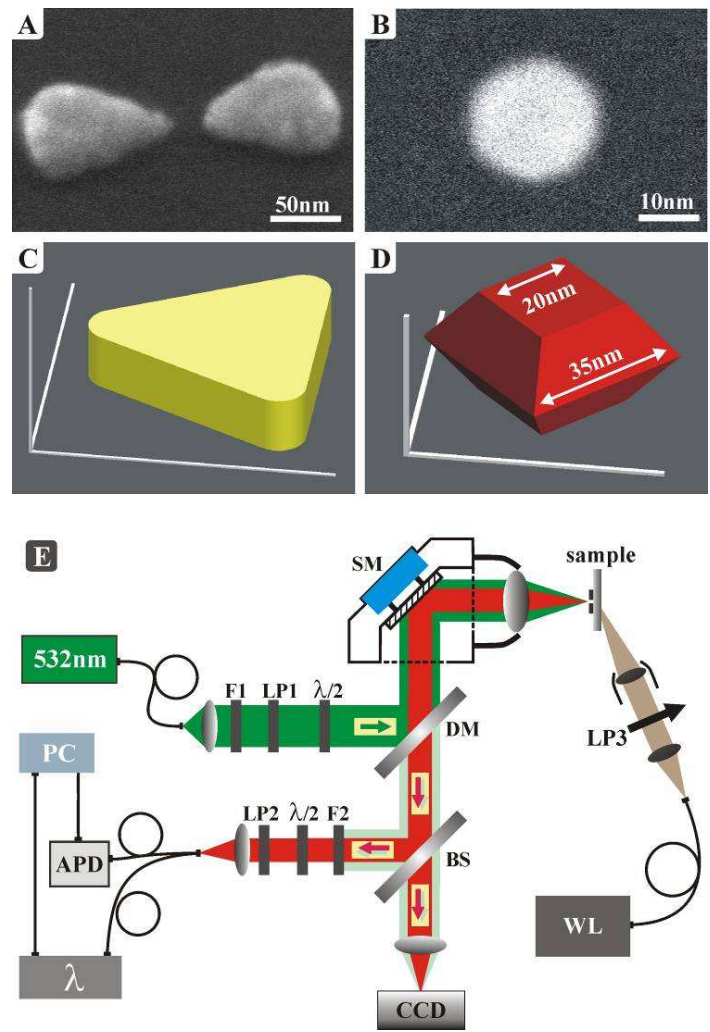

**Figure 1**. A – SEM image of a typical bowtie nanoantenna. B – SEM image of a typical diamond nanocrystal (NC). C – An illustration of a bowtie triangle used in the simulations (the white bars are 100 nm in length). D – An illustration of the diamond nanocrystal used in the simulations (the white bars are 50 nm in length). E – An illustration for the experimental setup used in the optical measurements. Here WL denotes the halogen lamp, SM a Princeton Instruments piezo-controlled scanning mirror, DM a dichroic mirror with a cut-off wavelength at 550 nm, and BS an uncoated BK7 glass beamsplitter. LP1,2 and 3 denote linear polarizers, F1 a 532 nm laser line filter, and F2 a 600 nm long pass filter. CCD denotes a camera used for imaging the sample surface, APD an avalanche photodiode photon counting module, and λ a Princeton instruments liquid nitrogen-cooled spectrometer, connected to a PC.

**Figure 2.**

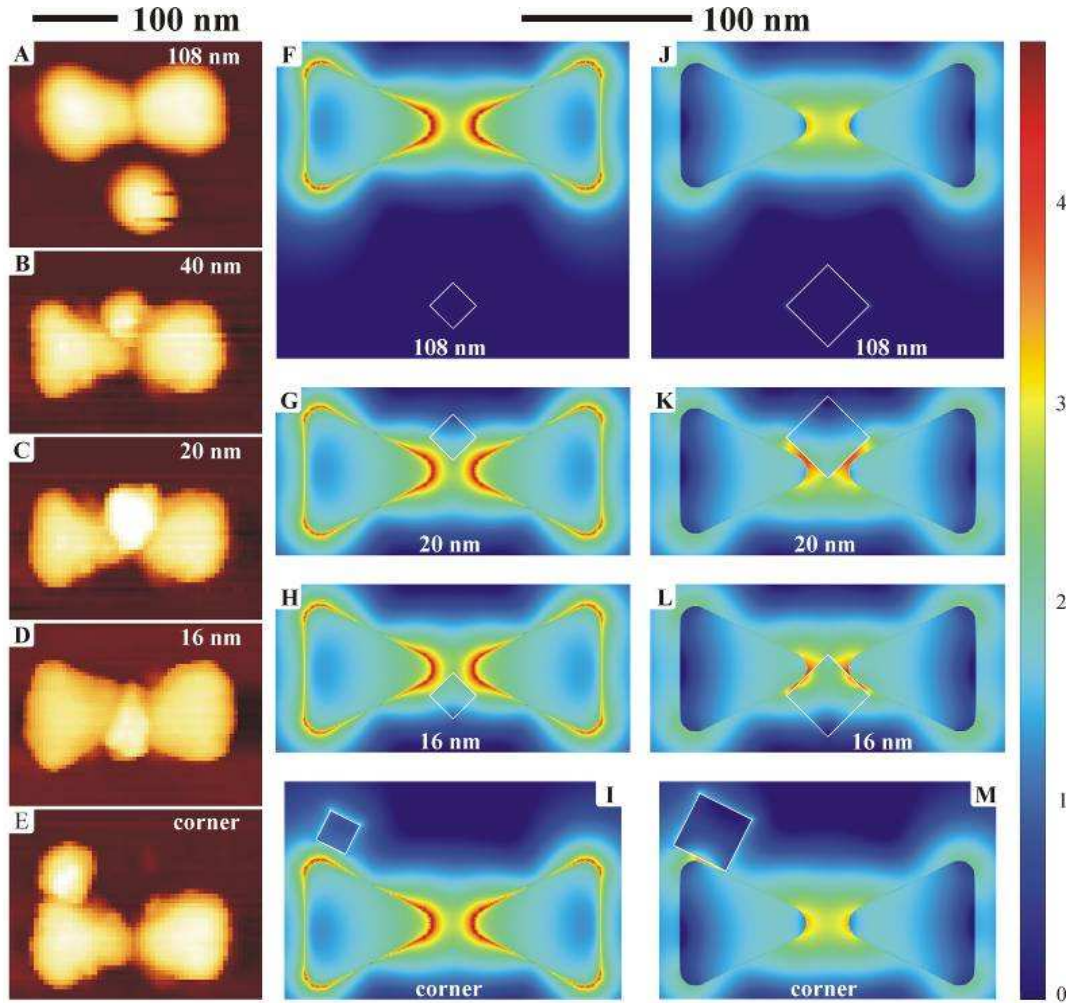

**Figure 2**. A-E – Atomic force microscopy images of the bowtie antenna and the diamond NC for five different NC locations. The measurements on each figure refer to the distance from the centre of the NC to the centre of the antenna gap. F-I – Field distributions around the BT-NC structure at the substrate level for the four of the five NC locations. J-M – Field distributions around the BT-NC structure at 17.5nm above the substrate level for four of the five NC locations. The 40 nm separation field distribution figures are omitted to save space, but are similar to the 108 nm figures, so present no additional information. The color bar on the right shows $|E|^2$ on a common log scale. The NC is outlined in white as a guide to the eye.

**Figure 3.**

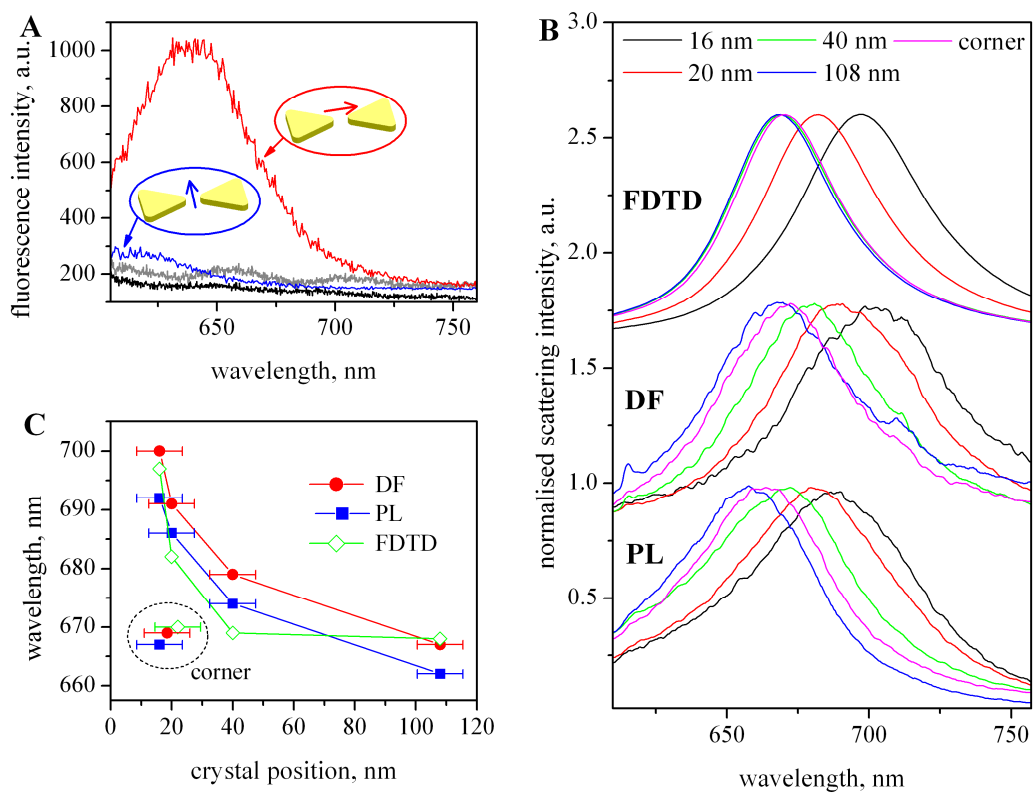

**Figure 3**. A – Polarization dependence of the gold luminescence spectrum. The red (blue) curve represents the component polarized parallel (orthogonal) to the main antenna axis. The black (gray) curve is the parallel (orthogonal) polarisation component of gold luminescence observed on an extended gold film of the same thickness. B – Normalized PL, DF and FDTD-simulated scattering spectra for the five NC locations in the vicinity of the BT discussed in Fig. 2. C – The plasmon resonance peak wavelength as a function of BT-NC distance, demonstrating the spectral shift. The peak wavelength when the NC is positioned next to one of the outer corners of the BT antenna shows almost no spectral shift, as expected, while the shift is significant for a NC in close proximity to the gap.